\documentclass[10pt,aps,prl,floatfix,showpacs,twocolumn,longbibliography,superscriptaddress,nofootinbib]{revtex4-1}
\usepackage[normalem]{ulem}
\usepackage{dcolumn}
\usepackage{bm}
\usepackage{amssymb}
\usepackage{amsmath}
\usepackage{graphicx}
\usepackage{amsfonts}
\usepackage{slashed}
\usepackage[colorlinks=true, linkcolor=blue, citecolor=blue, filecolor=blue,urlcolor=blue]{hyperref}
\usepackage[dvipsnames]{xcolor}
\usepackage{array}
\usepackage{multirow}
\allowdisplaybreaks
\usepackage{epsf}
\usepackage[nice]{nicefrac}
\usepackage{MnSymbol}
\usepackage{orcidlink}
\usepackage{physics}
\usepackage{tikz}
\usetikzlibrary{matrix,
fit,
calc,
backgrounds,
positioning,
arrows.meta,
decorations.pathreplacing,
shapes.arrows
}
\usepackage{float}
\usepackage{ulem}

\newcommand{\M}{\mathcal{M}}

\usepackage{xcolor}
\usepackage{pgf}

\makeatletter
\DeclareRobustCommand{\rainbow}[1]{\rainbow@s #1$\@sent$\@end}
\def\rainbow@s#1$#2$#3\@end{%
  \rainbow@words #1 \@nil
  \ifx\@sent#2\else
    \pgfmathsetmacro{\rainbowhue}{random(0,359)}%
    \textcolor[Hsb]{\rainbowhue,0.9,0.8}{$#2$}%
    \rainbow@s #3\@end
  \fi}
\def\rainbow@words#1 #2\@nil{%
  \rainbow@letters{#1}%
  \ifx\relax#2\relax\else\ \rainbow@words #2\@nil\fi}
\newcommand{\rainbow@letters}[1]{%
  \@tfor\@x:=#1\do{%
    \pgfmathsetmacro{\rainbowhue}{random(0,359)}%
    \textcolor[Hsb]{\rainbowhue,0.9,0.8}{\@x}%
  }%
}
\makeatother

\begin{document}

\title{Features of Non-Local Quantum Correlations in Nuclei}

\author{James W. T. Keeble\,\orcidlink{0000-0002-6248-929X}}
\email{jkeeble@physik.uni-bielefeld.de}
\affiliation{Fakult\"at f\"ur Physik, Universit\"at Bielefeld, D-33615, Bielefeld, Germany}

\author{Caroline E. P. Robin\,\orcidlink{0000-0001-5487-270X}}
\email{crobin@physik.uni-bielefeld.de}
\affiliation{Fakult\"at f\"ur Physik, Universit\"at Bielefeld, D-33615, Bielefeld, Germany}
\affiliation{GSI Helmholtzzentrum f\"ur Schwerionenforschung, Planckstra{\ss}e 1, 64291 Darmstadt, Germany}

\author{Federico Rocco\, \orcidlink{0009-0005-9826-012X}}
\email{frocco@physik.uni-bielefeld.de}
\affiliation{Fakult\"at f\"ur Physik, Universit\"at Bielefeld, D-33615, Bielefeld, Germany}

\author{Martin J.~Savage\,\orcidlink{0000-0001-6502-7106}}
\email{mjs5@uw.edu}
\thanks{On leave from the Institute for Nuclear Theory.}
\affiliation{InQubator for Quantum Simulation (IQuS), Department of Physics, University of Washington, Seattle, WA 98195}

\date{\today}

\begin{abstract}
Classical simulation of quantum many-body systems is challenged by the combined effects of entanglement and non-stabilizerness (magic). Non-local magic, $\mathcal{M}^{\mathrm{NL}}_\alpha$, quantifies the component of non-stabilizerness that cannot be removed by local unitary transformations and provides a measure of quantum complexity beyond entanglement alone. It is bounded from below by the antiflatness of the entanglement spectrum, $\mathcal{F}_A$. We calculate stabilizer R\'enyi entropies, non-local magic, entanglement-spectrum antiflatness, and entanglement entropy in shell-model ground states of more than 260 even-even, odd-even, and odd-odd nuclei, using a bipartition into proton and neutron sectors. We find that the previously established bound $4\mathcal{F}_A\leq\mathcal{M}^{\mathrm{NL}}_{\mathrm{lin}}$ holds for every nucleus studied and approaches saturation in the limit of separable states. Odd-odd nuclei exhibit markedly different patterns of entanglement and magic compared with even-even and odd-even systems, revealing distinctive signatures of proton-neutron quantum correlations and providing new insight into the complexity generated by nuclear interactions.
\end{abstract}

\maketitle

\textit{Introduction.}---
Whether a quantum many-body state can be represented and evolved on a classical computer is governed by its entanglement and non-stabilizerness (magic) structure, and identifying that structure is now central to deciding where quantum hardware is needed in simulations of nuclei and dense matter (for a recent review, see Ref.~\cite{Robin:2026lqp}).
The ground state wavefunctions of nuclei 
provide a well-suited setting in which to explore this question.
They are computationally demanding, strongly-interacting and highly-correlated systems, 
which 
carry substantial multipartite proton-neutron entanglement and non-stabilizerness~\cite{Brokemeier:2024lhq}, 
which makes them a unique setting in which general manifestations of quantum complexity can be investigated~\cite{Papenbrock:2003bj,Papenbrock:2003az,Papenbrock:2004jd,Legeza:2015fja,Kruppa:2020rfa,Robin:2020aeh,Stetcu:2021cbj,Kruppa:2021yqs,Johnson:2022mzk,Tichai:2022bxr,Gu:2023aoc,Gorton:2024hbb,Perez-Obiol:2023wdz,Kovacs:2024zoz,Tichai:2024cyd,Perez-Obiol:2024vjo,Brokemeier:2024lhq,Johnson:2025uoa,Shinde:2025xud,Shinde:2026hus,Robin:2026lqp}

Non-stabilizerness now stands on a resource-theoretic footing~\cite{Emerson:2013zse,Howard:2017maw}, 
with computable measures built from stabilizer R\'enyi entropies (SREs)~\cite{Leone:2021rzd,LeoneBittel2024},
$\M_\alpha$, that can be measured on quantum hardware~\cite{Oliviero2022}, evaluated for matrix product states~\cite{Haug2023} and sampled by Monte Carlo methods~\cite{Tarabunga2023,Brokemeier:2024lhq}.
These measures have been applied to an array of many-body systems, including lattice gauge theories, {\it e.g.}, Refs.~\cite{Tarabunga2023,Falcao:2024msg,Ebner:2025pdm,Froland:2026aff,Jha:2026ror,Jha:2026eni,Grieninger:2026bdq,Ebner:2026kwq,Minar:2026hqx},
nuclear and hypernuclear forces~\cite{Robin:2024oqc,Robin:2025dgw}, 
neutrino systems~\cite{Chernyshev:2024pqy,Hite:2026fsj}, 
top quarks~\cite{White:2024nuc,Yazgan_2026,CMS:2024zkc},
scattering in QED, QCD and gravity~\cite{Garrido:2025xpk,Robin:2025dgw,Gargalionis2025,Liu:2025qfl,Gargalionis:2026onv,Chu:2026yxm,Li:2026zum,Liu:2026gju}, 
and nuclei themselves~\cite{Brokemeier:2024lhq,Kadam:2026srm}.

The total non-stabilizerness, however, depends on the choice of basis and thus characterizes simulation costs only within that basis, rather than identifying an intrinsic need for quantum computing resources.
Non-stabilizerness that resides in a single subsystem can be removed by a local unitary, and states with large local non-stabilizerness but little entanglement are classically tractable.
Non-local non-stabilizerness (non-local magic)~\cite{Cao:2024nrx}
is invariant under local unitary transformations, and it has been argued that a smoothed version of it, 
the minimum over states within a small distance of the given state, 
bounds the classical hardness of simulating incompressible states~\cite{Cao:2024nrx}.
For arbitrary states, evaluating non-local non-stabilizerness is however difficult, because the minimization over local unitaries is not tractable beyond a few qubits. This has recently motivated the search for bounds derived from the entanglement spectrum.

For bipartitions, a close 
estimate of the non-local $\alpha=2$ SRE can be obtained from the Schmidt spectrum~\cite{Cao:2024nrx}.
The formulation involves constructing an auxiliary state with the same entanglement spectrum as the original state, whose total SRE provides a (tight) upper bound to the non-local SRE of the original state.
Numerical studies have found this bound to be saturated for systems of up to six qubits~\cite{Busoni:2026lvp,Robin:2026lqp}. Equality has been proved for $1\times N$ bipartitions~\cite{Franchini:2026dsg}, and is conjectured to hold for general bipartitions~\cite{Torre:2026pil,Liu:2026nah,Liu:2026twg}.
Exact formulations of non-local non-stabilizerness from the entanglement spectrum have also been recently demonstrated for the stabilizer fidelity~\cite{Sierant:2026qji,Viscardi:2026oos}.

On the other hand, it has been shown that a (loose) lower bound to the bipartite non-local non-stabilizerness is supplied by the antiflatness $\mathcal{F}_A$~\cite{Cao:2024nrx}, defined as the variance of the reduced density matrix~\cite{Tirrito:2023fnw}. 
In the two-qubit case, an exact equality between the non-local linear SRE $\M^{\mathrm{NL}}_{\mathrm{lin}}$ and antiflatness, $\M^{\mathrm{NL}}_{\mathrm{lin}} = 4\mathcal{F}_A$, has been numerically observed~\cite{Robin:2025dgw}, and now proven~\cite{Busoni:2026lvp}.
More recently, it has been proved~\cite{Liu:2026twg} 
that
\begin{equation}
    4\mathcal{F}_A \le \M^{\mathrm{NL}}_{\mathrm{lin}} 
    \ ,
    \label{eq:bound}
\end{equation}
for general bipartitions of $N$-qubit systems (see also Ref.~\cite{Jha:2026eni}).

In this study, we compute and compare the total and non-local non-stabilizerness, 
as well as the antiflatness of the entanglement spectrum
for the ground states of $> 260$ even-even, odd-even and odd-odd nuclei spanning the $p$-, $sd$-, $pf$- and $pf_{5/2}g$-shells, using a 
mapping of orbital occupations to 6-44 qubits, and a proton-neutron bipartitioning.
All nuclei satisfy the bound in Eq.~\eqref{eq:bound}, and the inequality is close to saturated for the most weakly entangled ones. 
Nuclei with odd numbers of both protons and neutrons are found to exhibit complexity structures that are quite distinct from those of even-even and odd-even nuclei.
Further,
non-stabilizerness and antiflatness appear to constitute indicators of shell closures, around which non-stabilizerness decreases while antiflatness tends to increase.
\\

\medskip
\textit{Method.}---
The nuclear wavefunction is defined as a linear combination of tensor products of proton $(A)$ and neutron $(B)$ Slater determinants,
\begin{equation}
    \ket{\Psi} = \sum_{i,j} C_{ij} \ket{\phi_i}_{A} \otimes \ket{\chi_j}_{B}
    \; .
    \label{eq:SM_wf}
\end{equation}
To make the computation of complexity measures tractable, 
the single-particle space is partitioned into a fully occupied inert core, an active valence space, and the remaining unoccupied orbitals. 
Such a formulation is known as the nuclear shell model~\cite{RevModPhys.77.427}, 
which provides an effective representation of the 
nuclear states~\footnote{Note that the nuclear shell model has been put in the context of {\it ab-initio} methods, see {\it e.g.} Refs.~\cite{Stroberg:2019mxo,10.3389/fphy.2020.00379}}. 
Within this framework, the valence nucleons interact via an empirical two-body Hamiltonian of the form,
\begin{equation}
    \hat{H} = \sum_p \varepsilon_p a^\dagger_p a_p + \frac{1}{4} \sum_{pqrs} \bar{v}_{pqrs} \, a^\dagger_p a^\dagger_q a_s a_r \; ,
    \label{eq:SM_Hami}
\end{equation}
where $a^\dagger_p$ ($a_p$) creates (annihilates) a nucleon in single-particle state (orbital) $p$, $\varepsilon_p$ is the corresponding single-particle energy, and $\bar{v}_{pqrs} = v_{pqrs} - v_{pqsr}$ are the antisymmetrized interaction matrix elements.
In this work, we employ the Cohen-Kurath~\cite{Cohen1965}, 
USDB~\cite{Brown2006} and GXPF1A~\cite{Honma2005} interactions 
for the $p$-, $sd$- and $pf$-shell respectively,
and explore a few isotopic chains going into the $pf_{5/2}g$-shell with the JUN45 interaction~\cite{Honma2009}.
These interactions are fit to reproduce sets of experimental nuclear properties, and thus incorporate effects of three-nucleon (and beyond) forces in a phenomenological way. They do not include the effects of isospin breaking or electromagnetism.
The ground-state wavefunctions are obtained using the {\tt BIGSTICK} code~\cite{Johnson2018,Johnson2013}, which employs the Lanczos algorithm, and are chosen to have the minimum value of angular momentum projection $J_z$~\footnote{That is, $J_z=0$ for nuclei with even numbers of nucleons, and $J_z=1/2$ for odd-nucleon systems. A resulting good value of the total angular momentum $J$ is ensured by the spherical symmetry of the interactions and model spaces.}.

The measures of quantum complexity computed below use a Jordan-Wigner-like mapping in which the occupation number of each nucleonic orbital is mapped onto a qubit. 
\footnote{Note that this is the same occupation-number encoding used in the {\tt BIGSTICK} code.} 
Each Slater determinant, 
$\ket{\phi_i}_{A} \otimes \ket{\chi_j}_{B}$ in Eq.~\eqref{eq:SM_wf}, 
is then represented by a computational basis state of $N_q=12$, $24$, $40$ and $44$ qubits for the $p$-, $sd$-, $pf$- and $pf_{5/2}g$-shell, respectively~\footnote{Nuclear states with only protons or only neutrons in the active space require $N_q/2$ qubits for their representation.}.

Following the natural bipartitioning of Eq.~\eqref{eq:SM_wf} into proton ($A$) and neutron ($B$) sectors, we compute the corresponding von Neuman entanglement entropy~\footnote{Note that proton-neutron entanglement entropy and the spectra of sets of shell model nuclei have been previously computed in {\it e.g.} Refs.~\cite{Johnson:2022mzk,Perez-Obiol:2023wdz,Shinde:2025xud,Shinde:2026hus} and \cite{Papenbrock:2003bj,Papenbrock:2003az}, respectively.}

\begin{equation}
    S = - \rm{Tr} \left( \rho_A \, log_2 \rho_A \right) \; ,
    \label{eq:entropy}
\end{equation}
and antiflatness of the entanglement spectrum~\cite{Tirrito:2023fnw}
\begin{equation}
    \mathcal{F}_A(\Psi) = \operatorname{Tr}\{\rho_A^3\}-\operatorname{Tr}^2\{\rho_A^2\} \; ,
    \label{eq:antiflatness}
\end{equation}
where $\rho_A=  \operatorname{Tr}_B \ket{\Psi}\bra{\Psi}$ is the reduced density matrix in the proton sector.
Non-stabilizerness is quantified with the rank-$\alpha$ Stabilizer R\'enyi Entropy ($\alpha$-SRE)~\cite{Leone:2021rzd},
\begin{equation}
\mathcal{M}_{\alpha}(\ket{\Psi})= -\log_2 d + \frac{1}{1-\alpha} \log_2
\left( \sum_{P} \Xi_P^{\alpha} \right) \; ,
\label{eq:Renyi_entropy_def1}
\end{equation}
where $P$ denotes the elements of the Pauli group for $N_q$ qubits (with phase $+1$), $d=2^{N_q}$ and $\Xi_P\equiv\frac{1}{d} \bra{\Psi}P\ket{\Psi}^2$.
In the present study, we use $\alpha=2$ and its linear form, $\M_{\mathrm{lin}} = 1 - 2^{-\mathcal{M}_2}$.
The practical computation of $\mathcal{M}_2 (\ket{\Psi})$ is described in the End Matter.

Non-local non-stabilizerness quantifies the minimum non-stabilizerness under local 
unitary transformations
$\hat{U}_A \otimes \hat{U}_B$ in the proton sector and neutron sector, separately.
As explicit minimization is intractable for most nuclei, the non-local non-stabilizerness is determined by employing the estimate  in which an auxiliary state is 
constructed from the Schmidt coefficients $\{ \lambda_k \}$ of $\ket{\Psi}$~\cite{Cao:2024nrx} as
\begin{equation}
    \left|\psi^{\prime}\right\rangle_{A B}=\sum_{k=0}^{r-1} \sqrt{\lambda_k}\left|s_k\right\rangle\left|s_k\right\rangle,
    \label{eq:schmidt_state}
\end{equation}
where $r$ is the Schmidt rank, 
the coefficients are ordered~\footnote{The non-stabilizerness of $\ket{\psi^{\prime}}$ depends on the ordering of the weights, and is smallest for descending order~\cite{Cao:2024nrx}. } $\lambda_0\ge\lambda_1\ge\dots\ge\lambda_{r-1}$, and $\{\ket{s_k}\}$ denote stabilizer states.
As stated above, $\mathcal{M}^{\mathrm{NL}}_2 (\ket{\Psi}) \approx \mathcal{M}_2(\ket{\psi^{\prime}})$, which we use in the following, along with the linear form $\M^{\mathrm{NL}}_{\mathrm{lin}} (\ket{\Psi}) = 1 - 2^{-\mathcal{M}^{\mathrm{NL}}_2 (\ket{\Psi})} \approx 1 - 2^{-\mathcal{M}_2 (\ket{\psi '})}$.
In practice we use the computational-basis states from Eq.~\eqref{eq:SM_wf} as a set of stabilizer states in Eq.~\eqref{eq:schmidt_state}, which ensures to preserve the relevant symmetries of the nuclear state (value of $J_z$)~\footnote{Using computational-basis states in Eq.~\eqref{eq:schmidt_state} has recently been shown to be optimal~\cite{Sierant:2026lvg}.}.

\begin{figure}[!ht]
    \centering
    \includegraphics[width=.9\columnwidth]{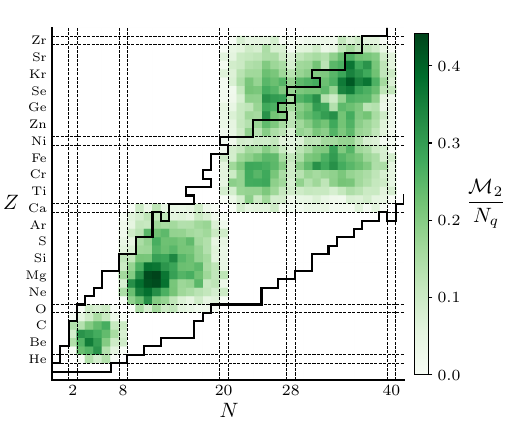}
    \caption{Total non-stabilizerness $\mathcal{M}_2 (\ket{\Psi})$ of the nuclear ground states, 
    divided by the number of mapping qubits $N_q$. The $x$ and $y$ axes denote numbers of neutrons ($N$) and protons ($Z$).  
    Dashed lines denote shell closures, the solid lines mark the experimentally determined driplines~\cite{IAEA_LiveChart}. The $N \leftrightarrow Z$ symmetry in the present calculations is due to omitted isospin-breaking effects.}
    \label{fig:chartm2}
\end{figure}

\medskip
\textit{Results.}---
Figure~\ref{fig:chartm2} shows the total non-stabilizerness $\mathcal{M}_2(\ket{\Psi})$ of the nuclear ground states, divided by the number of qubits $N_q$ used in the mapping, for nuclei in the $p$, $sd$ and $pf$ shells~\footnote{The $\alpha=2$ SREs of $p$- and $sd$-shell nuclei have previously been computed 
using wavefunctions with $J_z =J$~\cite{Brokemeier:2024lhq}.}.
The average value is approximately constant across the shells ($\approx 0.16 - 0.17$), showing that the non-stabilizerness grows linearly in the number of qubits (orbitals) in the present systems.
This is also shown in Fig.~\ref{fig:M2_vs_Nq} in the End Matter.
The non-stabilizerness per qubit however varies substantially across the nuclear chart, with distinct regions of enhancement, see also the End Matter. 
Interestingly, the largest values occur in intermediate-mass nuclei near $^{24}$Mg, for which $\M_2/N_q \simeq 0.44$, about half of the maximal non-stabilizerness values ($1 - 2/N_q$) reached by Haar random states~\cite{Leone:2021rzd}. 
As discussed in Ref.~\cite{Brokemeier:2024lhq}, 
these nuclei lie in a region characterized by large deformation.
In the $pf$-shell $^{70}$Br has the largest total non-stabilizerness, $\mathcal{M}_2$ = 16.45(7), corresponding to $\sim 0.41$ per qubit, while $^{66}$As is a local minimum in the top-right corner. 

\begin{figure}[!ht]
    \centering
    \includegraphics[width=.9\columnwidth]{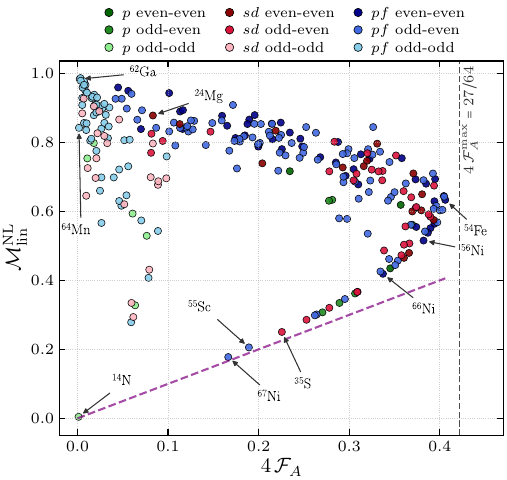}
    \caption{Non-local non-stabilizerness $\M^{\mathrm{NL}}_{\mathrm{lin}}(\ket{\Psi})$ against $4\mathcal{F}_A$ for the ground states of $>260$ nuclei, with the proton-neutron bipartition. 
    The purple dashed line is $\M^{\mathrm{NL}}_{\mathrm{lin}}=4\mathcal{F}_A$.
    The green, red, and blue colors correspond to nuclei of the $p$-, $sd$- and $pf$-shell, respectively.
    The shades distinguish nuclei with even and/or odd numbers of protons and neutrons.
    The vertical grey dashed line shows the maximal antiflatness for qubit systems~\cite{Jasser:2026hrg}.
    }
    \label{fig:af_vs_nl}
\end{figure}

Figure~\ref{fig:af_vs_nl} shows $\M^{\mathrm{NL}}_{\mathrm{lin}}$ versus $4\mathcal{F}_A$ for 
the same $>260$ nuclei. 
The non-local non-stabilizerness of each nucleus lies on or above the dashed-line corresponding to $\M^{\mathrm{NL}}_{\mathrm{lin}} = 4\mathcal{F}_A$
showing that the inequality in Eq.~\eqref{eq:bound} holds, 
across all considered valence spaces (mapped to 6 to 40 qubits).
Most nuclei lie far from that bound indicating that their non-local non-stabilizerness incorporates large effects beyond what is captured by antiflatness.
The figure for even-even and odd-even systems displays a ``boomerang'' structure. 
In the lower part of that structure lie nuclear states that are dominated by a single Schmidt value ($\lambda_0 \gtrsim 0.8$) and thus are close to product states across the proton-neutron cut.
Those with larger dominating Schmidt value have smaller antiflatness and nearly saturate the inequality $\M^{\mathrm{NL}}_{\mathrm{lin}} = 4\mathcal{F}_A$. 
Such nuclei typically exhibit a simple structure of a single particle or hole in one nucleon species coupled to a closed subshell of the other, which is seen in odd-even systems. For example, $^{67}$Ni ($Z$=28, $N$=39, $\lambda_0$=0.951) and $^{35}$S ($Z$=16, $N$=19, $\lambda_0$=0.928) are clean illustrations.
Even-even nuclei exhibit more complex structures and thus lower Schmidt values. For example, $^{66}$Ni ($Z$=28, $N$=38) has $\lambda_0$=0.861 while $^{56}$Ni ($Z$=$N$=28) has $\lambda_0$=0.827. 
As the entanglement spectrum 
develops a tail,
the non-local non-stabilizerness $\M^{\mathrm{NL}}_{\mathrm{lin}}$ starts to deviate from the lower bound $4\mathcal{F}_A$, while the antiflatness grows to reach a value near the maximum $4\mathcal{F}_A = 27/64 \approx 0.42$ for qubit systems~\cite{Jasser:2026hrg} in $^{54}$Fe and $^{54}$Ni.
In the upper branch of the ``boomerang'', the non-local non-stabilizerness 
increases as  the antiflatness decreases.

Odd-odd nuclei show distinct complexity structure, with 
$4\mathcal{F}_A\leq 0.1$, and $\M^{\mathrm{NL}}_{\mathrm{lin}} \gg 4\mathcal{F}_A$ for a majority of them.
A few odd-odd nuclei with $\M_{\rm{lin}}^{\rm NL} \lesssim 0.7$ possess two comparable, and large, leading Schmidt values.
Those are mainly one-particle or one-hole states in each sector, such as $^{14}$N, $^{54}$Sc or $^{34}$P. 
$^{14}$N, in particular, is essentially a perfect Bell (stabilizer) state in the $1p_{1/2}$ subshell with an entanglement entropy $S=1.01$.
The $^{78}$Y ground state shows a similar structure in $2p_{1/2}$, although it exhibits more mixing with other subshells, moving $^{78}$Y further from an ideal Bell state.

We note that rescaling the antiflatness by the square of the purity transforms the boomerang shape exhibited by even-even and even-odd nuclei to a near-universal function,
while leaving the odd-odd nuclei scattered above that curve, see Fig.~\ref{fig:nl_af_sqpurity} in the End Matter.
\\

\begin{figure}[!ht]
    \centering
    \includegraphics[width=.9\columnwidth]{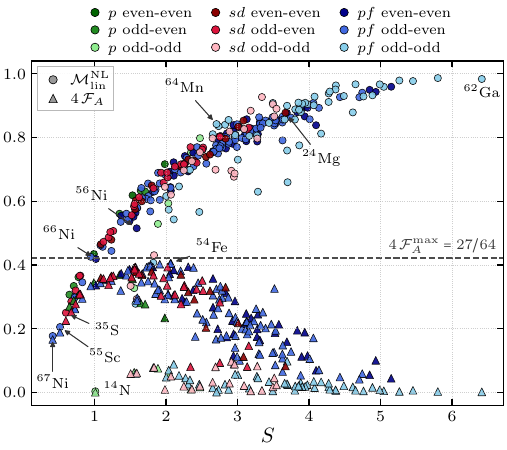}
    \caption{
    $\M^{\mathrm{NL}}_{\mathrm{lin}}$ and $4\mathcal{F}_A$, versus entanglement entropy $S$. The horizontal dashed line shows the maximal antiflatness reachable by qubit systems~\cite{Jasser:2026hrg}. }
    \label{fig:ratio_entropy}
\end{figure}

Figure~\ref{fig:ratio_entropy} shows  $\M^{\mathrm{NL}}_{\mathrm{lin}}$ and $4\mathcal{F}_A$ as a function of entanglement entropy $S$ across the bipartition.
At small $S$, the two quantities appear to 
approximately
coincide. 
As $S$ increases, $\M^{\mathrm{NL}}_{\mathrm{lin}}$ grows towards its maximum value, reaching about $0.98$ at $S\simeq6$, while $4\mathcal{F}_A$ separates into two branches: 
i) an upper branch 
containing the even-even and odd-even nuclei
that peaks near $0.4$ around $S\simeq 2$ and falls to $4\mathcal{F}_A \approx 0.1$ at $S \gtrsim 4.5$, 
and, ii) a lower branch 
containing the odd-odd nuclei
with $4\mathcal{F}_A\lesssim0.1$ at all $S$.
\\

\begin{figure}[!ht]
\centering
  \centering
  \includegraphics[width=.85\columnwidth]{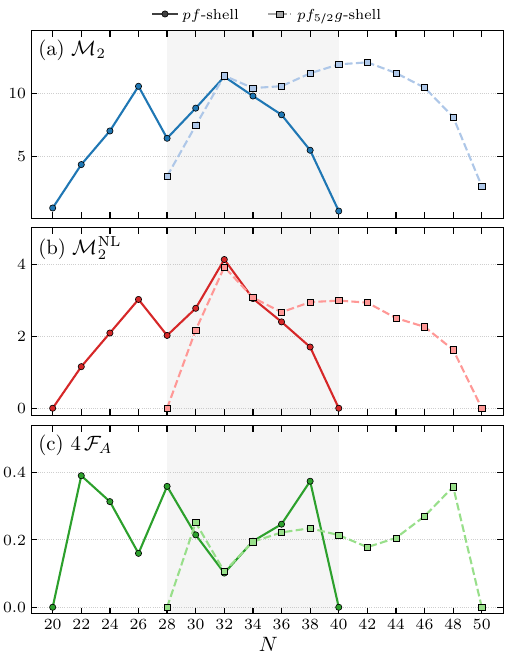}
      \caption{A comparison of measures of quantum complexity in the Germanium chain ($Z=32$) computed in the $pf$-shell (GXPF1A interaction~\cite{Honma2005}) and the  $pf_{5/2}g$-shell (JUN45 interaction~\cite{Honma2009}) model spaces.
      The upper panel shows the total non-stabilizerness $\M_2$ as a function of neutron number $N$, the middle panels shows the non-local non-stabilizerness $\M_2^{\rm NL}$, while the lower panel shows the antiflatness $4\mathcal{F}_A$.
      }
      \label{fig:pfgcomp}
\end{figure}
Finally, we study the evolution of the above measures of quantum complexity across the $N=40$ subshell closure.
Figure~\ref{fig:pfgcomp} shows $\M_2$, $\M_2^{\rm NL}$ and $\mathcal{F}_A$ in the Ge isotopic chain ($Z=32$), from $N=20$ to $N=50$, using the full $pf$-shell with the GXPF1A interaction~\cite{Honma2005} as above, and the $pf_{5/2}g$-shell with the JUN45 interaction~\cite{Honma2009}. 
The latter excludes the $1f_{7/2}$ subshell from the active space and instead includes the $1g_{9/2}$. 
Around $N=32$, where both model spaces describe the region constrained by experimental data, the calculated measures qualitatively agree.
At $N=28$, $\M_2$ and $\M_2^{\rm NL}$ display a drop with both interactions. Although this decrease arises from limitations of the model space in the JUN45 calculation, the (weaker) decrease observed with GXPF1A supports the interpretation of $N=28$ as a shell closure.
At $N=40$, while $\M_2$ and $\M_2^{\rm NL}$ decrease with GXPF1A because of the model-space limitation, they remain elevated when the $1g_{9/2}$ orbital is included in the $pf_{5/2}g$ space, consistent with the fact that $N=40$ is not a
full
shell closure.
In contrast to  measures of non-stabilizerness, 
the antiflatness appears to increase when approaching shell closures, 
as the
Ge chain descends along the upper branch of the ``boomerang'' 
in Fig.~\ref{fig:af_vs_nl}, only dropping due to active-space restrictions.
We observe similar behaviors in other isotopic chains with even proton numbers, such as Se, see the End Matter.

\medskip

\textit{Discussion and Outlook.}---
Nuclei provide a unique sandbox for exploring measures of quantum complexity.  
They are self-bound systems of two species of spin-1/2 fermions that are  nearly degenerate in mass,
arising from a finely-tuned strong force, exhibiting phenomena over a large range of energy scales.
These systems are emergent from the underlying quantum chromodynamics, a non-Abelian quantum field theory 
with massless gluons and light-quarks (compared to the confinement scale).
We have examined measures of quantum complexity in the ground state wavefunctions
of nuclei in the $p$-, $sd$- and $pf$-shells,
and explored a small number of isotopic chains into the $pf_{5/2}g$-shell.
Specifically, we have determined the total and non-local non-stabilizerness, 
and the antiflatness  of the entanglement spectrum.
All nuclei are found to satisfy the inequality, $4\mathcal{F}_A \le \M^{\mathrm{NL}}_{\mathrm{lin}}$.
Interestingly, odd-odd nuclei are seen to exhibit different structure(s) to the even-even and odd-even nuclei, in particular,
$4\mathcal{F}_A \lesssim 0.1$ and $4\mathcal{F}_A  \ll \M^{\mathrm{NL}}_{\mathrm{lin}}$ for the majority of them.
Even though even-even and odd-odd nuclei are known to differ in their structure properties from a phenomenological point of view, 
the substantial differences observed in their quantum complexity are striking.

Comparing the $pf$-shell GXPF1A interaction~\cite{Honma2005} with the $pf_{5/2}g$-shell JUN45 interaction~\cite{Honma2009} across the $N=28,40,50$ neutron shell gaps, shows that
(for the cases studied) 
full-state and non-local non-stabilizerness provide good indicators of shell closures
through their gradual decrease.
Antiflatness, however, can increase near a closure as the ground state moves along the upper branch of the boomerang plot toward its maximum, before decreasing as the state approaches a tensor-product state.

Going forward, it is important to understand the origin of the large differences between odd-odd and the other nuclei. 
A first step is to work with less structured nuclear forces, build in the various components piecewise, and recompute the complexity measures. Such a decomposition will likely impact the fine-tuning between components of the force, introducing spurious relations, but this remains to be determined.
Another line of exploration is to attempt to identify synthetic condensed matter systems that could possibly be engineered, to search for analogous behaviors in the quantum complexities.
Finally, translating the results of this work into a quantification of required quantum resources 
for preparing nuclear ground states on fault-tolerant quantum computers remains to be performed.
It is timely to carry out such determinations as quantum computers are approaching the point where early
simulations of nuclear dynamics are becoming possible, providing a starting point to compute processes of importance to experimental programs, and to reactor design sometime further into the future.
\\

\textit{Acknowledgments.}---
We would like to thank Florian Br\"okemeier  and S. Momme Hengstenberg for useful discussions. 
We would also like to acknowledge comments and discussions with participants of the third workshop on Many-Body Quantum Magic MBQM26, held at the European Center for Theoretical Physics ECT$^*$,
in particular S. Alterman, A. Hamma, and P. Love.
This work was supported, in part, 
by Universit\"at Bielefeld,  
by the Deutsche Forschungsgemeinschaft (DFG, German Research Foundation) through the CRC-TR 211 'Strong-interaction matter under extreme conditions'– project number 315477589 – TRR 211,
and by the MKW NRW under the funding code NW21-024-A.
This work was also supported, in part, by the U.S. Department of Energy, Office of Science, Office of Nuclear Physics, InQubator for Quantum Simulation (IQuS) under Award Number DOE (NP) Award DE-SC0020970 via the program on Quantum Horizons: QIS Research and Innovation for Nuclear Science, and through the Department of Physics and the College of Arts and Sciences at the University of Washington. 
This research used resources of the National Energy Research Scientific Computing Center, a DOE Office of Science User Facility supported by the Office of Science of the U.S. Department of Energy under Contract No. DE-AC02-05CH11231 using NERSC award NP-ERCAP0035809.
This work was enabled, in part, by the use of advanced computational, storage and networking infrastructure provided by the Hyak supercomputer system at the University of Washington.
Some of the computations in this work were also performed
on the GPU cluster at Bielefeld University. We
thank the Bielefeld HPC.NRW team for their support.

\bibliography{biblio_paper,biblio_paperv2}

%\clearpage
%\onecolumngrid
\section*{End Matter}

\textit{Computing non-stabilizerness via Monte Carlo sampling}---
The stabilizer 2-R\'enyi entropy, $\mathcal{M}_2 \equiv \mathcal{M}_2(\ket{\Psi})$ defined in Eq.~\eqref{eq:Renyi_entropy_def1}, can be evaluated with a replica representation in the space of computational basis states~\cite{Tarabunga2024}.
Under the JW mapping, the coefficients of Eq.~\eqref{eq:SM_wf} are the coefficients 
$c_\sigma\equiv C_{ij}$ 
of the state in the computational basis, 
$\ket{\Psi}=\sum_{\sigma}c_{\sigma}\ket{\sigma}$, where $\sigma$ is the occupation bit string of the Slater determinant $\ket{\phi_i}_{A} \otimes \ket{\chi_j}_{B}$. Then
\begin{eqnarray}
  &&   2^{-\mathcal{M}_2} = \sum\limits_{\sigma^{(1)}, \sigma^{(2)}, \sigma^{(3)}, \sigma^{(4)}}\left[c_{\sigma^{(1)}} c_{\sigma^{(2)}} c_{\sigma^{(3)}} c_{\sigma^{(1)}\oplus\sigma^{(2)}\oplus\sigma^{(3)}}\right. 
   \nonumber \\
    & & \ \ 
    \left.c_{\sigma^{(1)}\oplus\sigma^{(2)}\oplus\sigma^{(4)}}^* c_{\sigma^{(1)}\oplus\sigma^{(3)}\oplus\sigma^{(4)}}^* c_{\sigma^{(2)}\oplus\sigma^{(3)}\oplus\sigma^{(4)}}^* c_{\sigma^{(4)}}^*\right]\  
    ,  
\label{eq:M2_replica}
\end{eqnarray}
where $\oplus$ is the bitwise XOR of bit strings.

The direct summation of the replica trick of Eq.~\eqref{eq:M2_replica} scales as $\mathcal{O}(M^4)$ in the number of configurations, $M$, which for sparse wavefunctions, $M \ll 2^{N_q}$, scales more favorably than explicitly enumerating over all possible Pauli strings. 
In our case, we can exactly compute $\mathcal{M}_2$ for wavefunctions up to $M \approx 3 \times 10^4$ configurations 
by exploiting the real-valued character of the nuclear wavefunctions and permutation symmetry in Eq.~\eqref{eq:M2_replica}.
As the center of the $pf$-shell contains up to $2.3 \times 10^9$ configurations in the case of \textsuperscript{60}Zn ($J_z = 0$), we must pivot from a direct summation towards a statistical estimator for $\mathcal{M}_2$ that can approximate $\mathcal{M}_2$ to a sufficient degree.
We follow Ref.~\cite{sinibaldi2025non} to reformulate the summation of Eq.~\eqref{eq:M2_replica} as a statistical estimator in a 4-fold enlarged Hilbert space, $\mathcal{H}^{\otimes 4}$, with its associated wavefunction being a 4-fold tensor product of the original wavefunction, $\ket{\Phi} = \ket{\Psi}^{\otimes 4}$, with its basis state being $\ket{\eta} = \ket{\sigma^{(1)}} \otimes \ket{\sigma^{(2)}} \otimes \ket{\sigma^{(3)}} \otimes \ket{\sigma^{(4)}}$, where the superscript denotes the index of the replica. 
Within this setting, we can define a relation for the magic as follows,
\begin{align}
2^{-\mathcal{M}_2} = \frac{\langle \Phi \vert \hat{U} \vert \Phi \rangle}{\langle \Phi \vert \Phi \rangle} \approx \mathbb{E}_{\eta \sim \abs{\Phi}^2}\left[ \frac{\langle \eta \vert \hat{U} \vert \Phi \rangle}{\langle \eta \vert \Phi \rangle} \right],
\end{align}
where the action of $\hat{U}$ on a basis state, $\ket{\eta}$, is
\begin{align}
\begin{split}
\hat{U}|\eta\rangle = |&\sigma^{(2)}\oplus\sigma^{(3)}\oplus\sigma^{(4)}, \sigma^{(1)}\oplus\sigma^{(3)}\oplus\sigma^{(4)}, \nonumber \\
&\sigma^{(1)}\oplus\sigma^{(2)}\oplus\sigma^{(4)},\sigma^{(1)}\oplus\sigma^{(2)}\oplus\sigma^{(3)}\rangle,
\end{split}
\end{align}
and the term inside the statistical estimator can be viewed as a `local' contribution towards $\mathcal{M}_2$ for the configuration, $\ket{\eta}$.

To compute the statistical estimation of $2^{-\mathcal{M}_2}$, we must be able to efficiently sample configurations in proportion to the Born probability of the wavefunction in the enlarged Hilbert space, $\ket{\eta} \sim \abs{\Phi}^2$.
To do so, we use the Metropolis-Hastings sampling algorithm of Refs.~\cite{metropolis1949monte, hastings1970monte} in which our proposal distribution to navigate through the space of configurations, $\ket{\eta}$, is defined as an admixture of single- and double-orbital swaps that conserve $J_z$ with a 90\%\, -- 10\% weighting, respectively\footnote{As single-orbital swaps are confined to a single $J_z$ sector, double-orbital swaps are vital to ensure the ergodicity of our proposal distribution.}.
It should be noted that we take an absolute cutoff of 10\textsuperscript{-10} in the amplitudes of our wavefunctions~\footnote{Which is smaller than the adopted precision on the nuclear wavefunctions from {\tt BIGSTICK}.} 
when computing $\mathcal{M}_2$, and renormalize accordingly.

Our Markov Chain Monte Carlo (MCMC) sampling process consists of 2\textsuperscript{12} many-body walkers, which are evolved for chain lengths of 10\textsuperscript{5} steps.
To decorrelate our MCMC chains, we only record a `subset' of the chain by taking every t\textsuperscript{th} step, which depends on the particular shell at hand.
For example, the number of thinning steps between recorded results is taken to be 100$\times$ the number of orbitals in the respective shell; therefore, the $p$-, $sd$-, $pf$-, and the $pf_{5/2}g$-shells have 1200, 2400, 4000, and 4400 thinning steps, respectively.
In the case of some nuclei towards the center of the $pf$-shell, the thinning factor was taken to be 200$\times$ the number of orbitals, which totals 8000 thinning steps for some nuclei in the center of the $pf$-shell, e.g.\ \textsuperscript{60}Zn.
Therefore, the total chain length is defined as 10$\textsuperscript{5}$ multiplied by the respective thinning factor, which leads to a total chain length on the order of $10^8$ per MCMC walker and $\sim 5 \times 10^8$ samples per $\mathcal{M}_2$ evaluation.

% refactor
Once the MCMC sampling has completed, we can assess the accuracy of our MCMC sampling by the following metrics: the Monte Carlo Standard Error (MCSE), the potential scale reduction factor ($\hat{R}$), and the effective sample size (ESS) \cite{Vehtari2021}.
The MCSE is the standard error of the mean with the effective sample size rather than the total number of samples, which is used to account for the fact that a set of correlated samples contains 
an equivalent number of fewer \emph{i.i.d.}\ samples.
The potential scale reduction factor, $\hat{R}$, compares between-chain variance of means to the within-chain variance, approaching unity once the Markov chains have converged to the target distribution.
The effective sample size quantifies the number of \emph{i.i.d.}-equivalent draws per chain. We require the total samples across all chains to exceed 400$\times$ the number of Markov chains so that our statistical estimators have enough samples in order to have valid statistical estimates.
We use the `ranked' version of these metrics, which is defined as the maximum of the `bulk' and the `tail' of the distribution. The `tail' of the distribution is defined by samples below the 5\textsuperscript{th} percentile and above the 95\textsuperscript{th} percentile.
To compute $\mathcal{M}_2$, the MCMC code employs highly efficient CUDA kernels in CUDA.jl, 
and MCMC convergence is assessed using statistical diagnostics provided by the ArviZ.jl package~\cite{ARVIZ,Kumar2019}.
\\

\begin{figure}[!ht]
    \centering
    \includegraphics[width=.85\columnwidth]{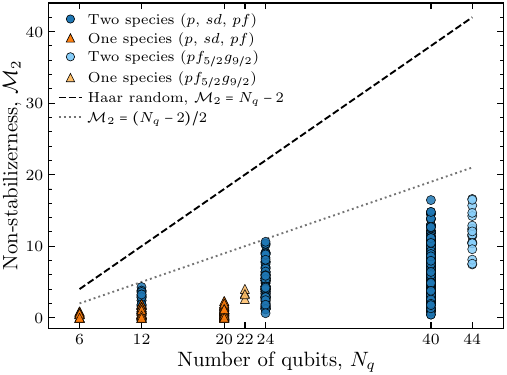}
    \caption{Non-stabilizerness $\M_2 (\ket{\Psi})$ of nuclear ground states in the $p$, $sd$, $pf$ and $pf_{5/2}g$ model spaces, as a function of number of qubits $N_q$ used for their mapping. Cases with only active protons or neutrons (one active specie) are shown with orange triangles, while those with two active species are shown with blue circles. The results in the $pf_{5/2}g$ shell ($N_q=22$ and $44$ qubits) include only the Ge and Se isotopic chains.
    A lower bound on the average value for Haar random states $\M_2 \approx N_q -2$~\cite{Leone:2021rzd} is shown with a thick dashed line, while the dotted line shows half of that value.}
    \label{fig:M2_vs_Nq}
\end{figure}

\textit{Scaling of non-stabilizerness}---
Figure~\ref{fig:M2_vs_Nq} shows the non-stabilizerness $\M_2 (\ket{\Psi})$ of nuclear ground states in all considered model spaces, as a function of number of qubits $N_q$ used for their mapping.
While $p$, $sd$ and $pf$ spaces span the full shells, the results for the $pf_{5/2}g$ space contain only Se and Ge isotopic chains.
It is clear that systems with two active species (neutrons and protons) display a large span of $\M_2$ values, with the largest ones scaling with $N_q$ and reaching approximately half the nearly maximal value characteristic of Haar-random states.
Nuclear states with only one active specie (neutrons or protons) display much less non-stabilizerness, that appears to scale only weakly with the number of qubits.
Whether the larger non-stabilizerness in nuclei arises solely from the presence of two fermion species or from the nature of the proton-neutron interaction remains to be explored.

\textit{Non-local non-stabilizerness and antiflatness of nuclei}---
\begin{figure}[!ht]
    \centering
    \includegraphics[width=0.9\columnwidth]{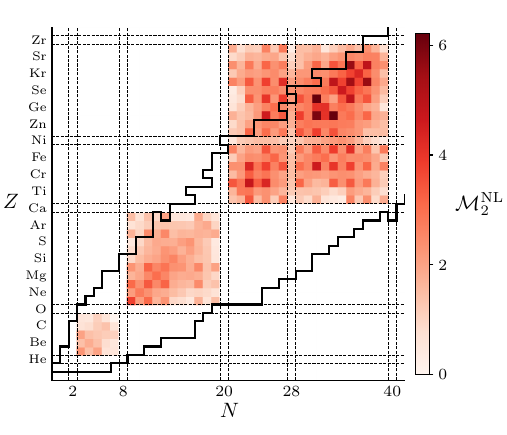}    \includegraphics[width=0.9\columnwidth]{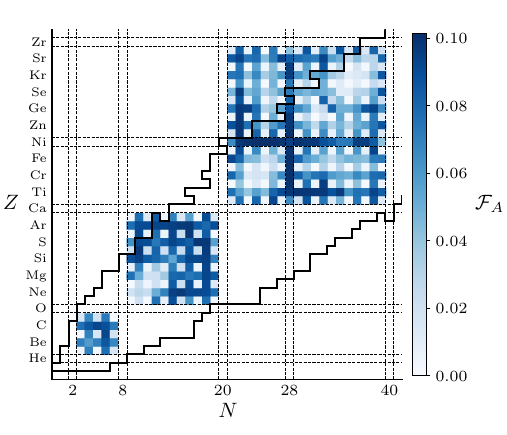}
    \caption{Non-local non-stabilizerness $\mathcal{M}^{\mathrm{NL}}_2$ (top) and antiflatness $\mathcal{F}_A$ (bottom) of shell-model ground states across the nuclear chart, for the proton-neutron bipartition. Dashed lines denote shell closures, the solid line marks the experimental driplines~\cite{IAEA_LiveChart}. }
    \label{fig:charts}
\end{figure}
Figure~\ref{fig:charts} shows the non-local non-stabilizerness $\mathcal{M}_2^{\rm{NL}}$ and the antiflatness for ground states 
of the $p$-, $sd$- and $pf$-shells, in the form of nuclear charts.
The two charts exhibit opposite checkerboard patterns, with odd-odd nuclei 
exhibiting high non-local non-stabilizerness and low antiflatness compared with even-even and odd-A nuclei.
In even-even nuclei, the entanglement spectrum is concentrated in a few dominant Schmidt weights, bringing the state in Eq.~\eqref{eq:schmidt_state} closer to a product across the cut and giving high antiflatness and low nonlocal magic. 
The unpaired proton and neutron in odd-odd systems, instead, couple their angular momenta over more configurations spanning the bipartition, distributing the entanglement spectrum over many comparable Schmidt weights. 
This flatter spectrum suppresses the antiflatness and results in a state far from both a product state and a maximally entangled state, with subsequently larger nonlocal magic.
\\

\begin{figure}[!ht]
\centering
  \centering
  \includegraphics[width=.85\columnwidth]{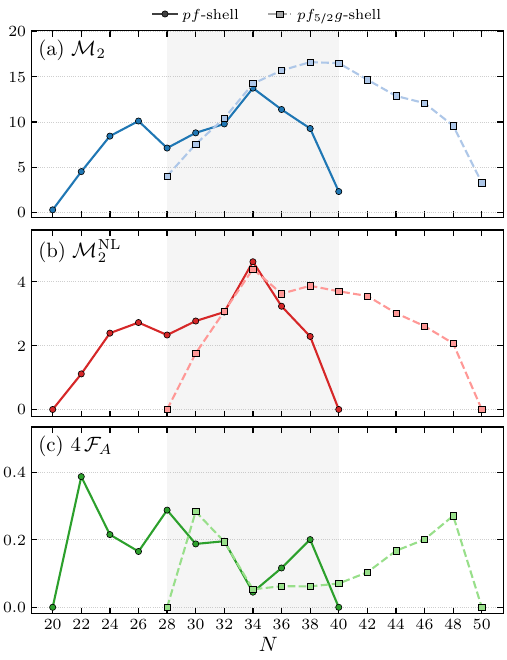}
      \caption{A comparison of measures of quantum complexity in the Selenium chain ($Z=34$) computed in the $pf$-shell (GXPF1A interaction~\cite{Honma2005}) and the  $pf_{5/2}g$-shell (JUN45 interaction~\cite{Honma2009}) model spaces.
      The upper panel shows the total non-stabilizerness $\M_2$ as a function of neutron number $N$, the middle panels shows the non-local non-stabilizerness $\M_2^{\rm NL}$, while the lower panel shows the antiflatness $4\mathcal{F}_A$.
      }
      \label{fig:pfgcompSe}
\end{figure}

\textit{Comparing the GXPF1A  and JUN45 in Selenium}---
Figure~\ref{fig:pfgcompSe} shows $\M_2$, $\M_2^{\rm NL}$ and $\mathcal{F}_A$ in the Se isotopic chain ($Z=34$), 
from $N=20$ to $N=50$, using the $pf$ valence space with the GXPF1A interaction~\cite{Honma2005} 
and the $pf_{5/2}g$ space with the JUN45 interaction~\cite{Honma2009}. 
The results show similar trends as those
obtained for the Ge chain, shown in Fig.~\ref{fig:pfgcomp} in the main text.
\\

\textit{Magic Without AntiFlatness}---   
Figures~\ref{fig:af_vs_nl} and \ref{fig:ratio_entropy} show that odd-odd nuclei typically exhibit large non-local magic despite small antiflatness. Although this may appear inconsistent at first glance, the two features are compatible, as we illustrate with a simple four-qubit example.
Consider a wavefunction
\begin{align}
|\psi\rangle & = \frac{1}{\sqrt{3}} \left[\ |0000\rangle + |0101\rangle + |1010\rangle \right]
\ ,
\end{align}
and the bi-partition into two 2-qubit regions.  
The reduced density matrix in one of the regions has eigenvalues $\frac{1}{3}, \frac{1}{3}, \frac{1}{3}, 0$,
yielding a vanishing antiflatness, but a non-local magic of $\M_2 = \log_2 \frac{9}{5}$.
The rank of the reduced density matrices is not of the form $2^k$, and hence the antiflatness is not indicative of the non-local magic~\footnote{We acknowledge discussions with Alioscia Hamma about this feature. 
See also discussions in Ref.~\cite{Ebner:2026kwq}.}. 
\\

\begin{figure}[!ht]
\centering
  \centering
  \includegraphics[width=.85\columnwidth]{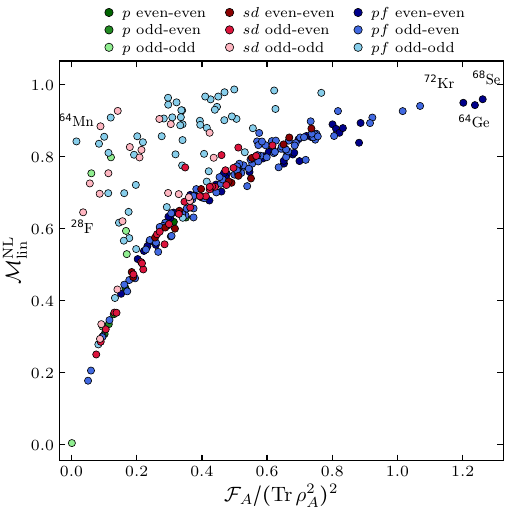}
      \caption{
      Non-local non-stabilizerness $\M^{\mathrm{NL}}_{\mathrm{lin}}(\ket{\Psi})$ against $\mathcal{F}_A/({\rm Tr}\rho_A^2)^2$ for the nuclear ground states shown in Fig.~\ref{fig:af_vs_nl},
      with the proton-neutron bipartition. 
    The green, red, and blue colors correspond to nuclei of the $p$-, $sd$- and $pf$-shell, respectively.
    The shades distinguish nuclei with even and/or odd numbers of protons and neutrons.
     }
      \label{fig:nl_af_sqpurity}
\end{figure}

\textit{Rescaling AntiFlatness by the Squared-Purity}---Figure~\ref{fig:nl_af_sqpurity} shows the non-local linear non-stabilizerness as a function of the antiflatness normalized to the square of the purity of the proton reduced density matrix $\rho_A$.
The boomerang shape exhibited by the even-even and even-odd nuclear transforms onto a nearly universal function, while the odd-odd nuclei appear to be somewhat arbitrarily distributed above that curve.
It is interesting to note that, when $\M^{\mathrm{NL}}_2(\ket{\Psi})$ is used, instead of $\M^{\mathrm{NL}}_{\mathrm{lin}}(\ket{\Psi})$, the universal curve for even-even and even-odd nuclei is essentially a straight line.

\end{document}